# Instabilities in Linear Accelerators

*M. Ferrario, M. Migliorati and L. Palumbo*
INFN-LNF and Universita' di Roma "La Sapienza"

Abstract

As a charged particle beam moves through perfectly conducting structures with varying cross-sectional boundaries - such as RF cavities, tapers, bellows, kickers, … - it induces both longitudinal and transverse electromagnetic fields called wakefields. In this lecture, we explore the fundamental characteristics of wakefields and illustrate key concepts using simple cases in cylindrical geometry, including a perfectly conducting pipe and the resonant modes of an RF cavity. Finally, we examine how wakefields influence beam dynamics in a LINAC, focusing on beam breakup instabilities and potential mitigation strategies.

## 1. Introduction

Self-induced electromagnetic (e.m.) fields, also known as wakefields, arise when a charged particle interacts with various vacuum chamber components. These components often have complex geometries, such as kickers, bellows, tapers, RF cavities, diagnostic elements, and other specialized devices. Analyzing these fields requires solving Maxwell's equations within the given structure, using the beam current as the field source. This is a highly complex task, necessitating the development of specialized computational tools capable of solving the electromagnetic problem in either the frequency or time domain. Several simulation codes, working with 2D or 3D geometries such as MAFIA, ABCI, and URMEL, have been widely used for accelerator design, as discussed in [1]. Nowadays new codes are also available, such as CST Microwave Studio [2], GdfidL [3], ACE3P [4], and Echo2D (3D) [5].

In this lecture, we first examine the fundamental properties of wakefields [6-14] and then illustrate key concepts through simple examples in cylindrical geometry, including the resonant modes of an RF cavity and a perfectly conducting pipe. In this last case, while space charge forces have been treated separately [15], these forces can also be considered a specific instance of wakefields, as demonstrated with an example in Appendix 2 [16].

Finally, we investigate the impact of wake fields on the transverse beam dynamics in a LINAC, focusing particularly on beam breakup instabilities. Assuming a high-energy beam where the longitudinal motion remains effectively "frozen," we outline a method to mitigate these instabilities [17].

## 2. Wakefields and Coupling Impedances

### 2.1 Wakefields

The e.m. fields created by a point charge interacting with the different devices of a particle accelerator act back on the charge itself and on any other charge of the beam. Let us focus our attention on the source charge $q_o$, and on the test charge $q$, assuming that both are moving with the same constant velocity $\mathbf{v}=\beta c$ on trajectories parallel to the axis.

Let $\mathbf{E}$ and $\mathbf{B}$ be the fields generated by $q_o$ inside a structure, $(s_o=vt, r_o)$ be the position of the source charge and $(s=s_o+z, r)$ be the position of the test charge $q$.

Since the velocity of both charges is along $\mathbf{z}$, the Lorentz force acting on $q$ has the following components:

$$\mathbf{F} = q\left[ E_z\hat{z} + \left(E_x - vB_y\right)\hat{x} + \left(E_y + vB_x\right)\hat{y} \right] \equiv \mathbf{F}_{//} + \mathbf{F}_{\perp}. \tag{1}$$

Thus, there can be two effects on the test charge $q$: a longitudinal force which changes its energy and a transverse force which deflects its trajectory. If we consider a device of length L, the energy gain [J] of the test charge is:

$$U = \int_0^L F_z \, ds \tag{2}$$

and the transverse deflecting kick [Nm] is:

$$\mathbf{M} = \int_0^L \mathbf{F}_{\perp} \, ds \ . \tag{3}$$

Note that the integration is performed over a given path of the trajectory. These quantities, normalised to the charges, are called **_wake-potentials_** (Volt/Coulomb) and are both functions of the distance $z$ between the source and the test charge:

*Longitudinal wake potential* $[V/C]$: $\qquad w_{//} = -\dfrac{U}{q_o q} \tag{4}$

*Transverse wake potential* $[V/Cm]$: $\qquad \mathbf{w}_{\perp} = \dfrac{1}{r_o}\dfrac{\mathbf{M}}{q_o q} \ . \tag{5}$

The minus sign in the longitudinal wake-potential means that the test charge loses energy when the wake is positive. Positive transverse wake means that the transverse force is defocusing.

As a first example, let us consider the longitudinal wake potential of the space charge. The longitudinal force inside a relativistic cylinder of radius $a$ travelling inside a cylindrical pipe of radius $b$ is given by [15]:

$$F_s(r,z) = \frac{-q}{4\pi\varepsilon_o\gamma^2}\left(1 - \frac{r^2}{a^2} + 2\ln\frac{b}{a}\right)\frac{\partial \lambda(z)}{\partial z} \ . \tag{6}$$

Note that since the space charge forces move together with the beam, and the electric field is constant along the beam pipe, we can derive the wake potential per unit length (Volt/Coulomb meter). To get the wake potential of a piece of pipe, we just multiply the result by the pipe length. Assuming $r \to 0$ and a charge line density given by $\lambda(z) = q_o\delta(z)$ we obtain:

$$\frac{dw_{//}(z)}{ds} = \frac{1}{4\pi\varepsilon_o\gamma^2}\left(1 + 2\ln\frac{b}{a}\right)\frac{\partial}{\partial z}\delta(z) \ . \tag{7}$$

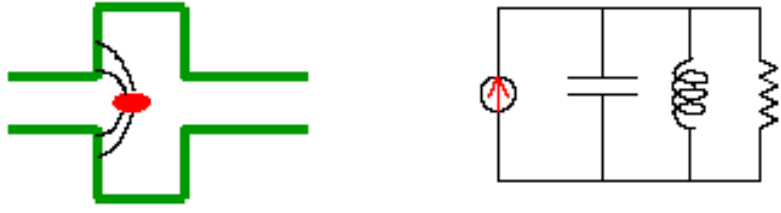

**Fig. 1:** Model of an RF cavity and the equivalent RLC parallel circuit driven by a current generator.

Another interesting case is represented by the longitudinal wake potential of a resonant mode in an RF cavity (it could be the fundamental mode or any higher-order mode – HOM) or a trapped mode in any machine device. When a charge crosses the resonant structure, the mode can be excited. To study its time evolution, we can treat it as an electric RLC circuit loaded by an impulsive current (a point charge), as shown in Fig. 1.

Just after the charge passage, the capacitor is charged with a voltage $V_o = V_c(0) = Cq_o$. This voltage then oscillates thanks to the combination of the capacitor and the inductance, each time reducing its amplitude (decaying) due to the resistance. The time evolution of this voltage is obtained by solving the circuit. In general, the differential equation for the voltage can be written as:

$$\ddot{V} + \frac{1}{RC}\dot{V} + \frac{1}{LC}V = \frac{1}{C}\dot{I} \ . \tag{8}$$

Since here we have an impulsive current, for t > 0 the voltage satisfies the following equation and boundary conditions:

$$\begin{aligned}
&\ddot{V} + \frac{1}{RC}\dot{V} + \frac{1}{LC}V = 0 \\
&V(t = 0^+) = \frac{q}{C} \equiv V_0 \\
&\dot{V}(t = 0^+) = \frac{\dot{q}}{C} = \frac{I(0^+)}{C} = \frac{V_0}{RC}
\end{aligned} \tag{9}$$

The solution of Eq. (9) is:

$$V(t) = V_0 e^{-\Gamma t}\left[\cos(\overline{\omega}t) - \frac{\Gamma}{\overline{\omega}}\sin(\overline{\omega}t)\right] \tag{10}$$

$$\overline{\omega}^2 = \omega_r^2 - \Gamma^2$$

where $\omega_r^2 = \frac{1}{LC}$ and $\Gamma = \frac{1}{2RC}$.

If a test charge follows the source at a distance $z = ct$ ($z$ is positive behind the charge), in passing through the device, it will gain an energy equal to $qV(z)$. By considering the definition of longitudinal wake potential given by Eq. (4), we obtain:

$$w_{//}(z) = \frac{-V(z)}{q_o} = w_o e^{-\Gamma z/c}\left[\cos(\overline{\omega}z/c) - \frac{\Gamma}{\overline{\omega}}\sin(\overline{\omega}z/c)\right], \tag{11}$$

which is represented in Fig. 2.

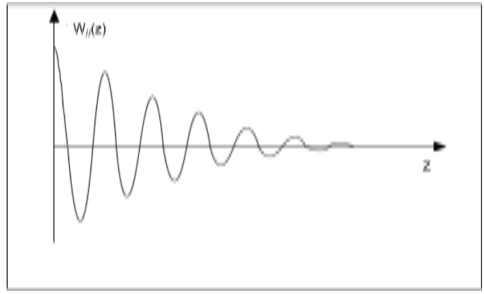

Fig. 2: Qualitative behavior of a resonant mode wake potential of a point charge.

## 2.2 Loss Factor

It is also useful to define the *loss factor* as the normalised energy lost by the source charge $q_o$:

$$k = -\frac{U(z=0)}{q_o^2} \qquad (12)$$

Although, in general, the loss factor is given by the longitudinal wake at $z = 0$, for charges travelling with the light velocity, the longitudinal wake potential is discontinuous in $z = 0$, as shown in Fig. 3.

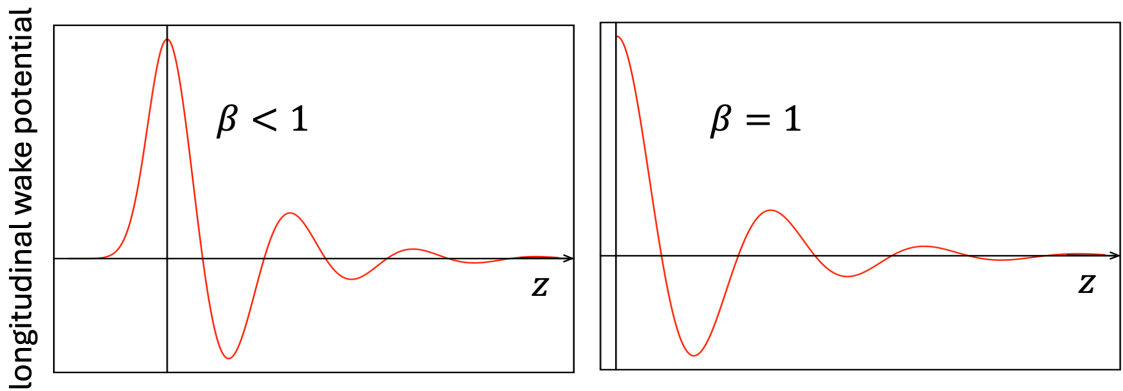

Fig. 3: Examples of longitudinal wake potentials: left $\beta < 1$, right $\beta = 1$.

The exact relationship between $k$ and $w(z \to 0)$ is given by the *beam loading theorem* [18].

## 2.3 Beam Loading Theorem

When the source charge travels with the speed of light $v = c$, it leaves the e.m. fields behind it, which is the reason why we call these fields "wakefields". Any e.m. perturbation produced by the charge cannot overtake the charge itself. This means that the longitudinal wake potential vanishes in the region $z < 0$. This property is a consequence of the *"causality principle"*. Causality requires that the longitudinal wake potential of a charge travelling with the velocity of light is discontinuous in the origin.

The beam loading theorem states that, in this case, the relation between the wake potential and the loss factor is:

$$k = \frac{w_{//}(z \to 0)}{2} \ . \tag{13}$$

For example, the beam loading theorem is fulfilled by the wake potential of the resonant mode. In fact, the energy lost by the source charge $q_0$ loading the capacitor is: $U = \frac{CV_o^2}{2} = \frac{q_o^2}{2C}$, thus giving $k = \frac{1}{2C}$, to compare with: $w_{//}(z \to 0) = \frac{1}{C}$.

## 2.4 Relationship between transverse and longitudinal forces

Another important feature worth mentioning is the differential relationship existing between longitudinal and transverse forces and, consequently, between the wake potentials:

$$\begin{aligned} \nabla_\perp F_{//} &= \frac{\partial}{\partial z} \mathbf{F}_\perp \\ \nabla_\perp w_{//} &= \frac{\partial}{\partial z} \mathbf{w}_\perp \end{aligned} \ . \tag{14}$$

The above relations are known as *"Panofsky-Wenzel theorem"* [19].

## 2.5 Coupling impedance

The wake potentials are used to study the beam dynamics in the time domain ($s=vt$). If we consider the equations of motion in the frequency domain, we need the Fourier transform of the wake potentials. This is particularly useful to study the collective effects in circular accelerators. Since these quantities have Ohms units, they are called *coupling impedances*.

*Longitudinal impedance* [$\Omega$]:
$$Z_{//}(\omega) = \frac{1}{v} \int_{-\infty}^{\infty} w_{//}(z) e^{-i\frac{\omega z}{v}} dz \tag{15}$$

*Transverse impedance* [$\Omega$/m]:
$$\mathbf{Z}_\perp(\omega) = \frac{i}{v} \int_{-\infty}^{\infty} \mathbf{w}_\perp(z) e^{-i\frac{\omega z}{v}} dz \ . \tag{16}$$

The longitudinal coupling impedance per unit of pipe length ($\Omega$/m) of the space charge wake potential given by Eq. (7) is:

$$\frac{\partial Z_{//}(\omega)}{\partial s} = \frac{1}{v} \int_{-\infty}^{\infty} \frac{\partial w_{//}(z)}{\partial s} e^{-i\frac{\omega z}{v}} dz = \frac{1 + 2\ln(b/a)}{v 4\pi\varepsilon_o \gamma^2} \int_{-\infty}^{\infty} \frac{d}{dz}\delta(z) e^{-i\frac{\omega z}{v}} dz \tag{17}$$

where: $\int_{-\infty}^{\infty} \delta'(z) f(z) dz = f'(0)$, so that:

$$\frac{\partial Z_{//}(\omega)}{\partial s} = \frac{-i\omega Z_o}{4\pi c \beta^2 \gamma^2} \left(1 + 2\ln\frac{b}{a}\right) \ . \tag{18}$$

The longitudinal coupling impedance of a resonant mode corresponding to the wake potential of Eq. (11) is given by:

$$Z_{//}(\omega) = \frac{R_s}{1 - iQ_r\left(\frac{\omega_r}{\omega} - \frac{\omega}{\omega_r}\right)}, \qquad (19)$$

where $R_s = \frac{w_{//}}{2\Gamma}$ is the shunt impedance and $Q_r = \frac{\omega_r}{2\Gamma}$ is the quality factor, quantities that we can obtain with electromagnetic computer codes. Note that the loss factor is also determined: $k = \frac{\omega_r R_s}{2Q_r}$.

Similarly, it is possible to define in the transverse plane the transverse wake potential and coupling impedance of a resonant mode, given, respectively, by

$$w_\perp(z) = \frac{\omega_r R_{s\perp}}{Q_r} \frac{c}{\bar{\omega}_\perp} e^{-\Gamma_\perp z/c} \sin(\bar{\omega}_\perp z/c)$$

$$Z_\perp(\omega) = \frac{c}{\omega} \frac{R_{s\perp}}{1 - iQ_r\left(\frac{\omega_r}{\omega} - \frac{\omega}{\omega_r}\right)} \qquad (20)$$

with $R_{s\perp} = \frac{R_s}{b^2}$ the transverse shunt impedance, and the other quantities defined similarly to the longitudinal plane.

## 2.6 Wake potential and energy loss of a bunched distribution

From the knowledge of the wake potential of a point charge, it is possible to evaluate the energy lost or gained by a single charge $e$ inside a bunch having longitudinal density λ(z) and total charge $q_{tot}$.

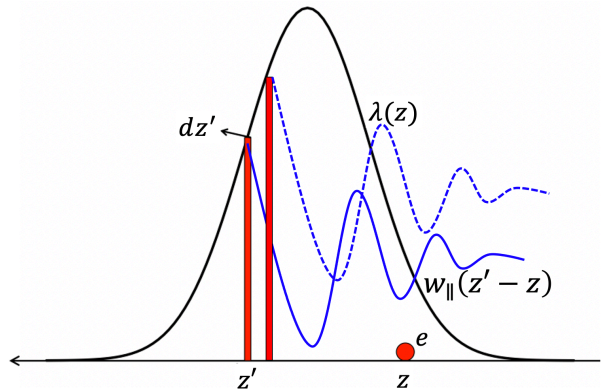

**Fig. 4:** Superposition principle to evaluate the energy lost by a charge in a bunch.

To this end, we use the superposition principle and calculate the effect on the charge by the whole bunch through the convolution integral, as illustrated in Fig. 4. The energy lost results equal to:

$$U(z) = -e \int_{-\infty}^{\infty} w_{//}(z'-z) \lambda(z') dz' \ . \tag{21}$$

This permits us to define the *wake potential of a distribution*:

$$W_{//}(z) = -\frac{U(z)}{eq_{tot}} \ . \tag{22}$$

The total energy lost by the bunch is computed by summing up the loss of all particles:

$$U_{bunch} = \frac{1}{e} \int_{-\infty}^{\infty} U(z') \lambda(z') dz' \ . \tag{23}$$

The wake potential defined by Eq. (22) depends on the particular charge distribution of the beam, and it is generally obtained as a result of electromagnetic codes. However, usually, the tools used to study collective effects, such as tracking codes, require, as input, the wake potential of a point charge (wakefield). It is, therefore, desirable to know what is the effect of a single charge (which is equivalent to finding the Green function of a structure) in order to reconstruct the fields produced by any charge distribution.

### 3. Wake Fields Effects in Linear Accelerators

#### 3.1 Energy Spread

The effect of the longitudinal wake potential is that of changing the energy of individual particles depending on their position in the bunch, thus resulting in an energy spread inside the beam.

For example, the energy spread induced by the space charge force per unit length in a Gaussian bunch is given by:

$$\frac{dU(z)}{ds} = -e \int_{-\infty}^{\infty} \frac{dw_{//}(z'-z)}{ds} \lambda(z') dz' = -\frac{eq}{4\pi\varepsilon_o \gamma^2 \sqrt{2\pi}\sigma_z^3} \left(1 + 2\ln\frac{b}{a}\right) z \, e^{-(z^2/2\sigma_z^2)} \ . \tag{24}$$

The bunch head ($z < 0$) gains energy, while the tail loses energy.

In a similar way, one can show that the energy loss induced by a resonant HOM on the charges inside a rectangular uniform bunch of length $l_0$, under the approximation that $\Gamma z/c \ll 1$ and $\bar{\omega}z/c \ll 1$, is given by:

$$U(z) = \frac{-eq_{tot}w_0}{2} \frac{\sin\left[\frac{\omega_r}{c}\left(\frac{l_o}{2} + z\right)\right]}{\left(\frac{\omega_r l_o}{2c}\right)} \ . \tag{25}$$

## 3.2 Beam Break Up

Due to any misalignments of the accelerating structures of a LINAC, a beam injected off-center executes betatron oscillations. The beam displacement produces a transverse wake potential in all the devices crossed during the flight, which can deflect the trailing charges as shown in Fig. 5, leading to an instability called single-bunch beam break-up [11]. It is important to note that there is also a multi-bunch beam break-up produced by the transverse wake potential of a bunch acting on another one that we do not treat in this lecture.

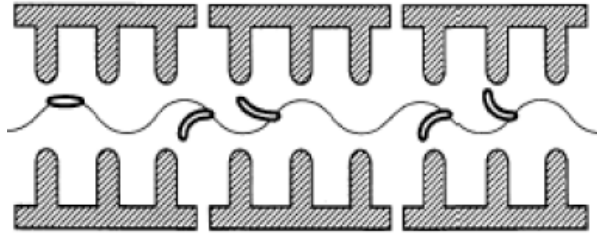

**Fig. 5:** Single-bunch beam break-up originated by an injection error [11].

In order to understand the effect, we consider, as a first example, a simple model with only two charges $q_1=Ne/2$ (leading = half bunch) and $q_2=e$ (trailing = single charge).

The leading charge executes free betatron oscillations:

$$y_1(s) = \hat{y}_1 \cos\left(\frac{\omega_y}{c} s\right). \tag{26}$$

The trailing charge, at a distance $z$ behind, over a length $L_w$ experiences a deflecting force proportional to the displacement $y_1$, and dependent on the distance $z$. According to the definition of transverse wake potential, we can write this force averaged over the length of the structure as:

$$\left\langle F_y^{self}(z,y_1) \right\rangle = \frac{Ne^2}{2L_w} w_\perp(z) y_1(s). \tag{27}$$

Notice that $L_w$ is the length of the device for which the transverse wake potential has been computed. For example, in the case of a cavity cell, $L_w$ is the length of the cell and $w_\perp(z)$ its wake potential. This force drives the motion of the trailing charge so that its equation of motion can be written as:

$$y_2'' + \left(\frac{\omega_y}{c}\right)^2 y_2 = \frac{Ne^2 w_\perp(z)}{2E_o L_w} \hat{y}_1 \cos\left(\frac{\omega_y}{c} s\right). \tag{28}$$

This is the typical equation of a resonator driven at the resonant frequency.

The solution is given by the superposition of the "free" oscillation and a "forced" oscillation which, being driven at the resonant frequency, grows linearly with $s$, as shown in Fig. 6.

$$y_2(s) = \hat{y}_2 \cos\left(\frac{\omega_y}{c} s\right) + y_2^{forced} \tag{29}$$

$$y_2^{forced} = \frac{cNe^2 w_\perp(z) s}{4\omega_y E_o L_w} \hat{y}_1 \sin\left(\frac{\omega_y}{c} s\right). \tag{30}$$

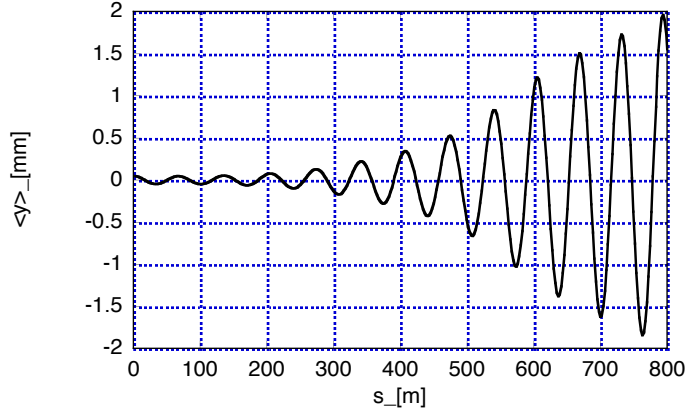

**Fig. 6:** HOMDYN [20] simulation of a typical BBU instability, 50 μm initial offset, no energy spread.

At the end of the LINAC of length $L_L$, the oscillation amplitude is grown by ($\hat{y}_1 = \hat{y}_2$):

$$\left(\frac{\Delta \hat{y}_2}{\hat{y}_2}\right)_{max} = \frac{cNew_\perp(z)L_L}{4\omega_y(E_o/e)L_w} . \qquad (31)$$

If the transverse wake is given per cell, the relative displacement of the tail with respect to the head of the bunch, depends on the number of cells. It also depends, of course, on the focusing strength through the frequency $\omega_y$. Observe also that the beam energy $E_0$ plays a beneficial role in the instability.

To extend the analysis to a particle distribution, we write the transverse equation of motion of a single particle with the inclusion of the transverse wakefield effects as [9,11]:

$$\frac{\partial}{\partial s}\left[\gamma(s)\frac{\partial y(z,s)}{\partial s}\right] + k_y^2(s)\gamma(s)y(z,s) = -\frac{e^2 N_p}{m_0 c^2 L_w}\int_z^\infty y(s,z')w_\perp(z'-z)\lambda(z')dz' , \qquad (32)$$

where $\gamma(s)$ is the relativistic parameter, $k_y(s)$ the betatron function, $N_p$ is the number of particles of the bunch and $\lambda(z)$ the longitudinal bunch distribution.

The solution of the above equation, or a similar one considering multiple bunches, has been found under different conditions, see, e.g. Refs. [21 - 23]. Here, for simplicity, we apply a perturbation method to obtain the solution at any order in the wakefield intensity. Indeed, we can write:

$$y(z,s) = \sum_n y^{(n)}(z,s) . \qquad (33)$$

The first-order solution is obtained from the equation without the contribution of the wake potential, that is

$$\frac{\partial}{\partial s}\left[\gamma(s)\frac{\partial y^{(0)}(z,s)}{\partial s}\right] + k_y^2(s)\gamma(s)y^{(0)}(z,s) = 0. \qquad (34)$$

It is important to notice that the above equation does not depend on $z$ anymore. This means that the bunch distribution remains constant along the structure.

If the s-dependence of $\gamma(s)$ and $k_y^2(s)\gamma(s)$ is moderate, we can obtain the solution of the above equation that represents the unperturbed transverse motion of the bunch. With the starting conditions $y(0) = y_m$ $y'(0) = 0$ we get:

$$y^{(0)}(s) = \sqrt{\frac{\gamma_0 k_{y0}}{\gamma(s)k_y(s)}} y_m \cos[\psi(s)] \tag{35}$$

with

$$\psi(s) = \int_0^s k_y(s')ds' . \tag{36}$$

The second-order differential equation is obtained by substituting the first-order solution in the rhs of Eq. (31) thus giving

$$\frac{\partial}{\partial s}\left[\gamma(s)\frac{\partial y^{(1)}(z,s)}{\partial s}\right] + k_y^2(s)\gamma(s)y^{(1)}(z,s) = -\frac{e^2 N_p}{m_0 c^2 L_w} y^{(0)}(s)\int_z^\infty w_\perp(z'-z)\lambda(z')dz' \tag{37}$$

We are interested in the forced solution of the above equation that can be written in the form

$$y^{(1)}(z,s) = -y_m \frac{e^2 N_p}{m_0 c^2 L_w} \sqrt{\frac{\gamma_0 k_{y0}}{\gamma(s)k_y(s)}} G(s)\int_z^\infty w_\perp(z'-z)\lambda(z')dz' , \tag{38}$$

where

$$G(s) = \int_0^s \frac{1}{\gamma(s')k_y(s')} \sin[\psi(s)-\psi(s')]\cos[\psi(s')]ds' =$$
$$= \frac{1}{2}\int_0^s \frac{\sin[\psi(s)-2\psi(s')]}{\gamma(s')k_y(s')}ds' + \frac{1}{2}\sin[\psi(s)]\int_0^s \frac{1}{\gamma(s')k_y(s')}ds' \tag{39}$$

The first integral undergoes several oscillations with s and, if $\gamma(s)$ and $k_y(s)$ do not vary much, it is negligible, so that we can finally write

$$y^{(1)}(z,s) = -y_m \frac{e^2 N_p}{2m_0 c^2 L_w} \sqrt{\frac{\gamma_0 k_{y0}}{\gamma(s)k_y(s)}} \sin[\psi(s)]\int_0^s \frac{ds'}{\gamma(s')k_y(s')}\int_z^\infty w_\perp(z'-z)\lambda(z')dz' . \tag{40}$$

The last integral represents the transverse wake potential produced by the whole bunch. This solution can then be substituted again in the rhs of Eq. (32) to obtain a third-order equation and so on. If we consider constant $\gamma(s)$ and $k_y(s)$, we obtain the same result of the two-particle model if we substitute $\lambda(z)$ with 1/2 in the particle positions.

As an example, in Fig. 7 we have represented a Gaussian bunch with the corresponding transverse wake potential produced in a cell of a Ka-band accelerating structure. For a conservative estimation of the BBU effect in this case one should use the maximum value of the wake potential within the bunch, thus eliminating the z-dependence.

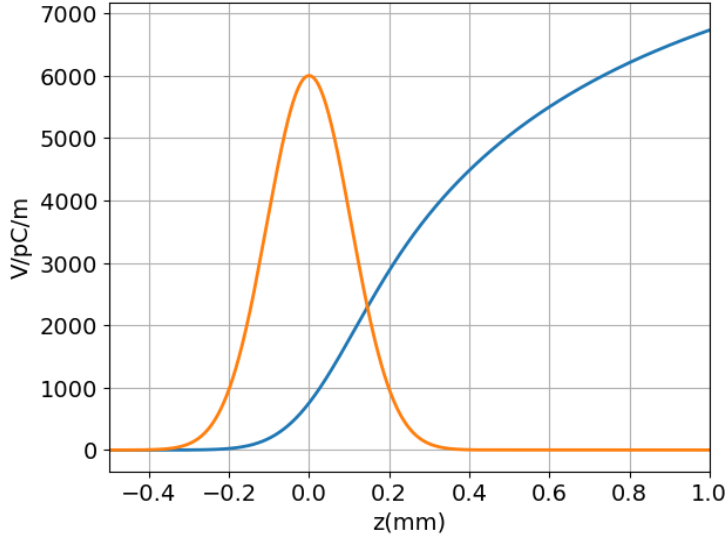

**Fig.7:** Transverse wake potential used to evaluate the single-bunch beam break-up.

If the BBU effect is strong, it is necessary to include higher-order terms in the perturbation expansion. Under the assumption of:

- rectangular bunch distribution $\lambda(z) = 1/l_0$, $-l_0/2 \leq z \leq l_0/2$, with $l_0$ the bunch length
- monoenergetic beam,
- constant acceleration gradient $dE_0/ds = const$,
- constant beta function,
- linear wake function inside the bunch $w_\perp(z) = w_{\perp 0} z/l_0$,

the sum of Eq. (33) can be written in terms of powers of the adimensional parameter $\eta$ also called BBU strength

$$\eta = \frac{e^2 N_p}{k_y (dE_0/ds)} \frac{w_{\perp 0}}{L_w} \ln\left(\frac{\gamma_f}{\gamma_i}\right) \tag{41}$$

with $\gamma_i$ and $\gamma_f$ respectively the initial and final relativistic energy parameters.

By using the method of the steeping descents [9], it is possible to obtain the asymptotic expression of $y(z,s)$ thus finding, at the end of the linac,

$$y(L_L) = y_m \sqrt{\frac{\gamma_i}{6\pi\gamma_f}} \eta^{-1/6} exp\left[\frac{3\sqrt{3}}{4}\eta^{1/3}\right] cos\left[k_y L_L - \frac{3}{4}\eta^{1/3} + \frac{\pi}{12}\right] \tag{42}$$

that, differently from the two-particle model and from the first-order solution, gives a tail displacement growing exponentially with $\eta$, resulting in a better agreement with the simulation shown in Fig 6.

## 3.3 BNS damping

According to Eq. (31), the BBU instability is mitigated at higher energies and also a strong focusing, increasing the betatron frequency $\omega_y$, helps to stabilize the beam. Notwithstanding, the BBU can be quite harmful and hard to take under control even after careful injection and steering.

A simple method to cure this instability has been proposed, observing that the high oscillation amplitude of the bunch tail is driven by the resonance condition with respect to the head. If the tail and the head move with a different frequency, this effect can be significantly removed [17], as we can see in Fig. 8 compared with Fig. 6.

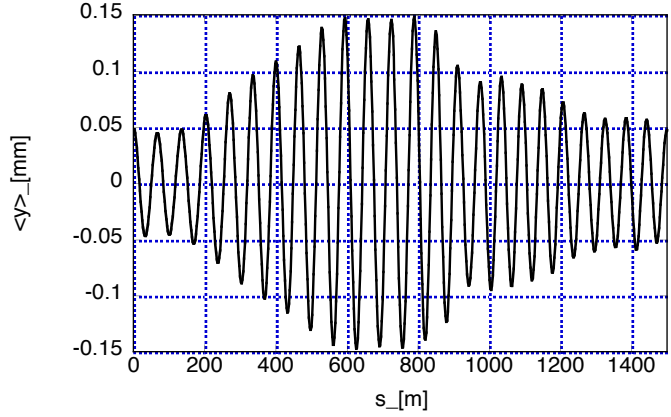

**Fig. 8:** HOMDYN simulation of a typical BNS damping, 50 μm initial offset, 2% energy spread.

Let us assume that the tail oscillates with a frequency $\omega_y + \Delta\omega_y$. Its equation of motion in the two-particle model, instead of Eq. (28), reads now:

$$y_2'' + \left(\frac{\omega_y + \Delta\omega_y}{c}\right)^2 y_2 = \frac{Ne^2 w_\perp(z)}{2E_0 L_w} \hat{y}_1 \cos\left(\frac{\omega_y}{c} s\right) \tag{43}$$

the solution of which is:

$$y_2(s) = \hat{y}_1 \cos\left(\frac{\omega_y + \Delta\omega_y}{c} s\right) + \frac{c^2 N e^2 w_\perp(z)}{4 \omega_y \Delta\omega_y E_0 L_w} \hat{y}_1 \left[\cos\left(\frac{\omega_y}{c} s\right) - \cos\left(\frac{\omega_y + \Delta\omega_y}{c} s\right)\right] . \tag{44}$$

The result is a free oscillation (first term) plus a beating between the two oscillations with a limited amplitude that does not increase any more with time.

Furthermore, by a suitable choice of $\Delta\omega_y$, it is possible to fully suppress the oscillations of the tail. Indeed, by setting:

$$\Delta\omega_y = \frac{c^2 N e^2 w_\perp(z)}{4 \omega_y E_o L_w} \tag{45}$$

we get:

$$y_2(s) = \hat{y}_1 \cos\left(\frac{\omega_y}{c} s\right) = y_1(s) . \tag{46}$$

The extra focusing $\Delta\omega_y$ at the tail can be obtained: 1) by using a RFQ, where the head and the tail see a different focusing strength, 2) by exploiting the energy spread across the bunch, which, because of the chromaticity, induces a spread in the betatron frequency. An energy spread correlated with the position is attainable with the external accelerating voltage, or with the wakefields.

## 4. Landau Damping

There is an important, natural stabilising effect against the collective instabilities called "Landau Damping" [9]. The basic mechanism relies on the fact that if the particles in the beam have a spread in their oscillation frequencies (synchrotron or betatron), their motion can't be coherent for a long time.

### 4.1 Driven oscillators

In order to understand the physical nature of this effect, we consider a simple harmonic oscillator, at rest for t < 0, driven by a harmonic force for t > 0.

$$\frac{d^2x}{dt^2} + \omega^2 x = A\cos(\Omega t) \;. \tag{47}$$

The general solution is given by the superposition of the free and forced oscillations, as shown in Appendix 1:

$$x(t) = \frac{A}{\omega^2 - \Omega^2}\left[\cos(\Omega t) - \cos(\omega t)\right] \;. \tag{48}$$

Let us assume now that the external force is driving a particle population characterised by a spread of natural frequencies of oscillation around a mean value $\omega_x$. Furthermore, let the forcing frequency $\Omega$ be inside the spectrum and such that $\delta \equiv \Omega - \omega \ll \omega_x$.

The motion of a given particle in the bunch can be approximated by:

$$x(t) \cong \frac{At}{2\omega_x}\sin(\omega_x t)\frac{\sin\left(\frac{\delta}{2}t\right)}{\left(\frac{\delta}{2}t\right)} \;. \tag{49}$$

Let us consider two particles in the bunch, one with $\delta=0$, and the other with $\delta\neq 0$. Both are at rest, and at t=0, and they start to oscillate with the same amplitude and phase (coherency). However, while the amplitude of the former charge grows indefinitely (driven at resonance), the latter reaches a maximum amplitude (beating of the two close frequencies of Eq. (48)). We say that the particle system has lost the coherency at the time when the beating amplitude is maximum, i.e. for t=$\pi/\delta$ .

We can also say that at any time t*, only those oscillators inside the bandwidth $|\delta| < \pi/$ t*, oscillate coherently. The longer we wait, the narrower is this coherent bandwidth, and, therefore, the less is the number of "coherent" particles.

### 4.2 Amplitude of oscillations

At any t*, we can divide the bunch population into two groups: the coherent particles, oscillating all together with an amplitude growing linearly with time, and the "incoherent" particles which have different phases and a saturated amplitude of oscillation.

It is interesting to see that, although the amplitude of the coherent oscillators grows linearly with time, the average amplitude of the whole system remains bounded. The reason is that the number of coherent particles decreases inversely with time. Indeed, the average amplitude oscillation of the whole system with $N$ particles can be written as

$$\langle x(t) \rangle^{max} = \frac{1}{N}\left[\sum_{coh} x(t) + \sum_{incoh} x(t)\right]^{max} . \tag{50}$$

The average of the displacements of the incoherent particles, being uncorrelated, is zero. We consider the time when the coherent particles have the maximum amplitude, so that we have:

$$\langle x(t) \rangle^{max} = \frac{N_{coh}}{N} x^{max}(t) = \frac{N_{coh}}{N} \frac{A}{2\omega_x} t . \tag{51}$$

On the other hand, the number of oscillators keeping the coherency decreases with time, by supposing that the frequency distribution of the particles is uniform in a frequency range $\Delta\omega$, we have:

$$N_{coh} = \frac{N}{\Delta\omega}\frac{\pi}{t} \quad \Rightarrow \quad \langle x(t) \rangle^{max} = \frac{\pi}{\Delta\omega}\frac{A}{2\omega_x} . \tag{52}$$

## 4.3  Energy of the system

What happens to the energy of the system? Also, in this case we distinguish the coherent and incoherent particles. The energy of the coherent particles grows quadratically with time, while the energy of the incoherent particles is bounded. In this case, although the number of coherent oscillators decreases with time, the total energy still grows linearly.

$$E(t) = E_{coh}(t) + E_{incoh}(t) \tag{53}$$

$$E_{coh}(t) = N_{coh}\left[\frac{1}{2}kx_{coh}^2(t)\right] \propto \frac{\pi}{2}\frac{NA^2}{\omega_x^2}\frac{1}{2\Delta\omega}t . \tag{54}$$

In conclusion, when a force drives such a system, only at the beginning we find the whole system following the external force. Afterwards, fewer and fewer particles are driven at the resonance. The result is that, although the system absorbs energy, the average amplitude remains bounded.

This mechanism works also when the driving force is produced by the bunch itself. In order to make the coherent instability to start, the rise time of the instability has to be shorter than the "de-coherency" time of the bunch.

## Appendix 1 – Driven Oscillators

Consider a harmonic oscillator with natural frequency ω and with an external excitation at frequency Ω:

$$\ddot{x} + \omega^2 x = A\cos(\Omega t).$$

The general solution is:

$$x(t) = x^{free}(t) + x^{driven}(t)$$
$$\cos(\Omega t) \Rightarrow e^{i\Omega t}$$
$$x^{free}(t) = \tilde{x}_m^f e^{i\omega t}$$
$$x^{driven}(t) = \tilde{x}_m^d e^{i\Omega t}$$

The driven solution (steady state) is found by direct substitution in the differential equation:

$$(\omega^2 - \Omega^2)\tilde{x}_m^d e^{i\Omega t} = A e^{i\Omega t} \quad \Rightarrow \quad x^{driven}(t) = \frac{A}{(\omega^2 - \Omega^2)} e^{i\Omega t}.$$

The general solution has to satisfy the initial condition at t=0. In our case we assume that the oscillator is at rest for t=0:

$$x^{free}(t=0) = -x^{driven}(t=0)$$
$$\tilde{x}_m^f = -\frac{A}{\omega^2 - \Omega^2}$$

thus, we get:

$$x(t) = \frac{A}{\omega^2 - \Omega^2}\left[e^{i\Omega t} - e^{i\omega t}\right]$$

taking only the real part:

$$x(t) = \frac{A}{\omega^2 - \Omega^2}\left[\cos(\Omega t) - \cos(\omega t)\right].$$

This expression is suitable for deriving the response of the oscillator driven at resonance or at a very close frequency: $\omega = \Omega + \delta$, with $\delta \to 0$. Defining: $\overline{\omega} = (\omega + \Omega)/2$, equivalent to $\omega = \overline{\omega} + \delta/2$ or $\Omega = \overline{\omega} - \delta/2$ the solution is given by:

$$x(t) = \frac{A}{2\overline{\omega}\delta} \left\{ \left[\cos(\overline{\omega}t)\cos(\delta t/2) + \sin(\overline{\omega}t)\sin(\delta t/2)\right] + \right.$$
$$\left. - \left[\cos(\overline{\omega}t)\cos(\delta t/2) + \sin(\overline{\omega}t)\sin(\delta t/2)\right] \right\}$$

that is:

$$x(t) = \frac{A}{\overline{\omega}\delta}\sin(\overline{\omega}t)\sin(\delta t/2) \equiv \frac{At}{2\overline{\omega}}\sin(\overline{\omega}t)\frac{\sin(\delta t/2)}{\delta t/2}$$

with the limit: $\quad \lim_{\delta \to 0} x(t) = \frac{At}{2\overline{\omega}}\sin(\overline{\omega}t).$

## Appendix 2 – Power Radiated by a Bunch passing through a Taper

In the case of a charge distribution, and with $\gamma \to \infty$, the electric field lines are perpendicular to the direction of motion and travel together with the charge [24], as shown in Fig. 9. In other words, the field map does not change during the charge flight, as long as the trajectory is parallel to the pipe axis. Under this condition, the transverse field intensity can be computed like in the static case, applying the Gauss and Ampere laws:

$$\int_S \varepsilon_o \mathbf{E} \cdot \mathbf{n} \, dS = \int_V \rho \, dV, \qquad \oint \mathbf{B} \cdot d\mathbf{l} = \mu_o \int_S \mathbf{J} \cdot \mathbf{n} \, dS. \qquad (A2.1)$$

Let us consider a cylindrical beam of radius $a$ and current $I$, with uniform charge density $\rho = \dfrac{I}{\pi a^2 v}$ and current density $J = \dfrac{I}{\pi a^2}$, propagating with relativistic speed $v = \beta c$ along the z-axis of a cylindrical perfectly conducting pipe of radius b, as shown in Fig. 9.

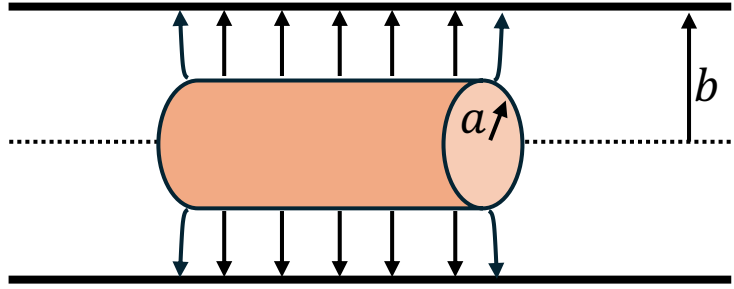

**Fig. 9:** Cylindrical bunch of radius $a$ propagating inside a cylindrical perfectly conducting pipe of radius $b$.

By applying the relations (A2.1), for the radial component of the electric field, one can obtain:

$$E_r = \frac{I}{2\pi\varepsilon_o a^2 v} r \quad \text{for} \quad r \leq a$$

$$E_r = \frac{I}{2\pi\varepsilon_o v} \frac{1}{r} \quad \text{for} \quad r > a \quad .$$

For the azimuthal magnetic field, the relation $B_\vartheta = \dfrac{\beta}{c} E_r$ holds.

The electrostatic potential satisfying the boundary condition $\varphi(b) = 0$ is given by:

$$\varphi(r,z) = \int_r^b E_r(r',z) dr' = \begin{cases} \dfrac{I}{4\pi\varepsilon_o v}\left(1 + 2\ln\dfrac{b}{a} - \dfrac{r^2}{a^2}\right) & \text{for} \quad r \leq a \\ \dfrac{I}{2\pi\varepsilon_o v} \ln\dfrac{b}{r} & \text{for} \quad a \leq r \leq b \end{cases}.$$

How can a perturbation of the boundary conditions affect the beam dynamics? Let's consider the following example: a smooth transition of length L (as a taper) from a beam pipe of radius b to a larger beam pipe of radius d is experienced by the beam [10]. To satisfy the boundary condition of a perfectly conducting pipe also in the tapered region, the field lines are bent as shown in Fig. 10. Therefore, there must be a longitudinal $E_z(r,z)$ field component in the transition region.

A test particle running outside the beam charge distribution $(r > a)$ will experience along the transition of length L a voltage given by [16]:

$$V = -\int_{z}^{z+L} E_z(r,z')dz' = -(\varphi(r,z+L) - \varphi(r,z)) = -\frac{I}{2\pi\varepsilon_o v} \ln\frac{d}{b}$$

that is decelerating if $d > b$. In order to sustain this induced voltage, the beam must lose the power given by:

$$P_{lost} = VI = \frac{I^2}{2\pi\varepsilon_o v} \ln\frac{d}{b} \, . \tag{A2.2}$$

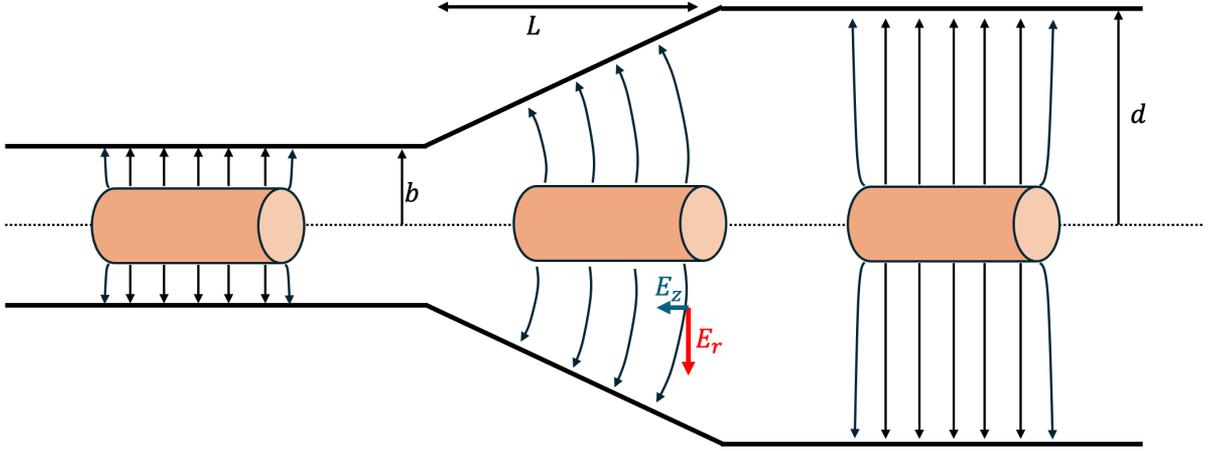

**Fig. 10:** Smooth transition of length L (taper) from a beam pipe of radius $b$ to a larger beam pipe of radius $d$.

When $d > b$, some energy must then be deposited by the beam in the fields: moving from left to right of the transition, the beam induces some fields in the additional room around the bunch (i.e. in the region $b < r < d$, $0 < z < L$) at the expense of only the energy available from the source, that is the kinetic energy of the beam itself.

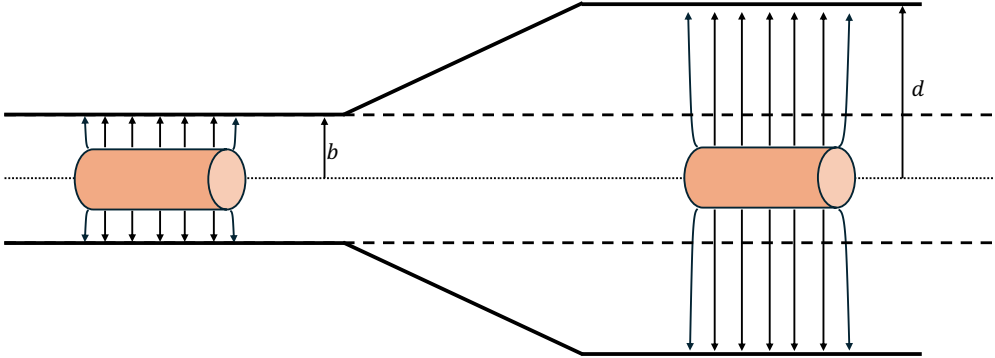

**Fig. 11:** During the beam propagation in the taper, additional em power flow is required to fill up the additional available room.

To verify such an interpretation, let us compute the electromagnetic power radiated by the beam to fill up the additional room available between $d$ and $b$, see Fig 11. Integrating the Poynting vector through the surface $\Delta S = \pi(d^2 - b^2)$ representing the additional power passing through the right part of the beam pipe [16], one obtains:

$$P_{em} = \int_{\Delta S}\left(\frac{1}{\mu}\vec{E}\times\vec{B}\right)\cdot\hat{n}dS = \int_{b}^{d}\frac{E_r B_\vartheta}{\mu} 2\pi r\,dr = \frac{I^2}{2\pi\varepsilon_o v}\ln\frac{d}{b}$$

that is exactly the same expression as Eq. (A2.2). Notice that if $d < b$, the beam gains energy. If $d \to \infty$ the power goes to infinity. Such an unphysical result is nevertheless consistent with the initial assumption of an infinite energy beam ($\gamma \to \infty$).